\documentclass[aps,preprint]{revtex4}%
\usepackage{amsfonts}
\usepackage{amsmath}
\usepackage{amssymb}
\usepackage{graphicx}%
\providecommand{\U}[1]{\protect\rule{.1in}{.1in}}

\begin{document}
\preprint{ }
\title{Time translation of quantum properties}
\author{Roberto Laura}
\affiliation{Facultad de Ciencias Exactas, Ingenier\'{\i}a y Agrimensura (UNR) and
Instituto de F\'{\i}sica Rosario (CONICET - UNR), Av. Pellegrini 250, 2000
Rosario, Argentina}
\email{rlaura@fceia.unr.edu.ar}
\author{Leonardo Vanni}
\affiliation{Instituto de Astronom\'{\i}a y F\'{\i}sica del Espacio (UBA - CONICET).
Casilla de Correos 67, Sucursal 28, 1428 Buenos Aires, Argentina.}
\email{lv@iafe.uba.ar}
\begin{abstract}
Based on the notion of time translation, we develop a formalism to deal with
the logic of quantum properties at different times. In our formalism it is
possible to enlarge the usual notion of context to include composed properties
involving properties at different times. We compare our results with the
theory of consistent histories.

\end{abstract}
\date{July 2008}
\maketitle

\section{Introduction.}

The absence of determinism is one of the main differences of the quantum
theory with the classical one. Nevertheless, it is necessary in quantum theory
to deal in a consistent way with expressions involving different properties of
the system at different times. For example, it is necessary to relate a
property of a microscopic system at a given time, previous to a measurement,
with a property of an instrument when the measurement is finished. Moreover,
in the famous double slit experiment it is argued about the impossibility to
say which slit a particle passed before producing a spot on a photographic
plate \cite{Fey}.

In a series of papers starting in 1984, an approach to quantum interpretation
known as consistent histories, or decoherent histories, has been introduced by
R. Griffiths \cite{Gri}, R. Omn\`{e}s \cite{Omn}, M. Gell-Mann and J. Hartle
\cite{G y H}. In their approach, the notion of history is defined as a
sequence of properties at different times. The probability for a history is
also defined through a formula motivated by the path integral formalism, but
with no direct relation with the Born rule. A consistency condition should be
satisfied by the possible different histories which can be included in a
legitimate description of the system, therefore the name of consistent
histories. For a given physical system the possible sets of consistent
histories depend on the state. This is not entirely satisfactory, because in
axiomatic theories of quantum mechanics the state is usually considered as a
functional on the space of observables, and it appears after these observables
in a somehow subordinate position. The importance of the notion of state
functionals acting on a previously defined space of observables was stressed
by one of us in references \cite{RL1} and \cite{RL2}.

In this paper we explore a different approach to define the probability of a
conjunction of properties at different times, and to discriminate which are
the properties that can be simultaneously considered in a description of the
system. Essentially, we consider the time translation of a property, already
at our disposal in the dynamic generated by the Schr\"{o}dinger equation. When
properties corresponding to different times are translated to a common single
time, we can apply to them the usual rules of compatibility between different
observables, and compute the probabilities using the Born rule.

In section II we present a brief summary of the theory of consistent
histories, and we discuss its application to a spin system. In section III we
give a short review of the notions of quantum logic, contexts and
probabilities. The notion of time translation of quantum properties is
introduced in section IV, where we also obtain the non distributive lattice of
time dependent properties. This definitions are used in section V to implement
expressions involving the conjunction of properties at different times, and to
define the compatibility of these type of expressions. Distributive lattices
of time dependent properties, called generalized contexts, are also obtained
in this section. The generalized contexts are compared with the theory of
consistent histories in section VI. The conclusions are given in section VII.

\section{The theory of consistent histories.}

In what follows we present the main features of the theory of consistent
histories, following essentially the approach given by R. Omn\`{e}s
\cite{Om3}, \cite{Om2}, \cite{Om4}, and we also discuss the application of
this theory to the case of a spin system.

Let us consider a state of a system at time $t_{0}$, represented by the state
operator $\widehat{\rho}_{t_{0}}$. An observable represented by an operator
$\widehat{A}_{j}$ with spectrum $\sigma_{j}$ is considered for each time
$t_{j}$ ($j=1,...,n$) of a sequence verifying $t_{0}<t_{1}<.....<t_{n}$. Each
spectrum $\sigma_{j}$ is partitioned by a family $\{\Delta_{j}^{(\mu)}\}$ of
mutually exclusive sets ($\cup_{\mu}\Delta_{j}^{(\mu)}=\sigma_{j}$,
$\Delta_{j}^{(\mu)}\cap\Delta_{j}^{(\mu^{\prime})}=\varnothing$ $(\mu\neq
\mu^{\prime})$).

The operator $\widehat{E}_{j}^{(\mu)}$ is the projector onto the subspace of
the Hilbert space corresponding to the subset $\Delta_{j}^{(\mu)}$ of the
spectrum $\sigma_{j}$, and it represents the property "the value of the
observable $A_{j}$ is in the set $\Delta_{j}^{(\mu)}$ at time $t_{j}$". These
projectors satisfy the equations $\widehat{E}_{j}^{(\mu)}\widehat{E}_{j}%
^{(\mu^{\prime})}=\delta_{\mu\mu^{\prime}}\widehat{E}_{j}^{(\mu^{\prime})}$
and $\sum_{\mu}\widehat{E}_{j}^{(\mu)}=\widehat{I}$.

A \textit{history} $a$ is defined by the property "(the value of $A_{1}$ is in
$\Delta_{1}^{(k_{1})}$ at $t_{1}$) and (the value of $A_{2}$ is in $\Delta
_{2}^{(k_{2})}$ at $t_{2}$) and ..... and (the value of $A_{n}$ is in
$\Delta_{n}^{(k_{n})}$ at $t_{n}$)". It is represented by the \textit{history
operator}%
\begin{equation}
\widehat{C}_{a}=\widehat{E}_{n}^{(k_{n})}(t_{n}).....\widehat{E}_{2}^{(k_{2}%
)}(t_{2})\,\widehat{E}_{1}^{(k_{1})}(t_{1}),\qquad\widehat{E}_{j}^{(k_{j}%
)}(t_{j})=e^{i\widehat{H}(t_{j}-t_{0})/\hbar}\widehat{E}_{j}^{(k_{j}%
)}e^{-i\widehat{H}(t_{j}-t_{0})/\hbar}, \label{ch1}%
\end{equation}
where $\widehat{H}$ is the Hamiltonian operator of the system.

The probability of the history $a$ is defined by the expression%
\begin{equation}
\Pr(a)=Tr(\widehat{C}_{a}\,\widehat{\rho}_{t_{0}}\,\widehat{C}_{a}^{\dag}).
\label{ch2}%
\end{equation}

As the probability should verify positivity, normalization and additivity, the
possible histories to be included in a valid description of the system should
verify some \textit{consistency conditions}. Sufficient conditions, given by
Gell-Mann and Hartle, are%
\begin{equation}
Tr(\widehat{C}_{a}\,\widehat{\rho}_{t_{0}}\,\widehat{C}_{b}^{\dag})=0,\qquad
a\neq b. \label{SC}%
\end{equation}

The theory of consistent histories is a framework suitable to include
properties of a system at different times in the language of quantum theory.
Moreover, these properties at different times are given a well defined
probability by Eq. (\ref{ch2}), provided we consider properties within a
consistent family of histories. Each family of consistent histories generate a
possible \textit{universe of discourse} about a quantum system. In general, it
is not possible to include two families in a single larger one. Different sets
of consistent histories are considered \textit{complementary descriptions} of
the system.

For simplicity we are going to consider the case $n=2$, involving histories
with only two different times $t_{1}$ and $t_{2}$. For the time $t_{1}$ the
spectrum of the observable $\widehat{A}_{1}$ is partitioned by two
complementary sets $\Delta_{1}$ and $\overline{\Delta}_{1}$ with projectors
$\widehat{E}_{1}$ and $\widehat{\overline{E}}_{1}$, while for the time $t_{2}$
the spectrum of the corresponding observable $\widehat{A}_{2}$ is partitioned
by the sets $\Delta_{2}$ and $\overline{\Delta}_{2}$ with projectors
$\widehat{E}_{2}$ and $\widehat{\overline{E}}_{2}$.

For this special case the necessary and sufficient consistency condition to
obtain well defined probabilities is%
\begin{equation}
\operatorname{Re}\,Tr\,(\widehat{E}_{1}(t_{1})\,\widehat{\rho}_{t_{0}%
}\,\widehat{\overline{E}}_{1}(t_{1})\,\widehat{E}_{2}(t_{2}))=0, \label{ch3}%
\end{equation}
which is called the Griffiths condition.

In this case, the sufficient Gell-Mann and Hartle condition is%
\begin{equation}
Tr\,(\widehat{E}_{1}(t_{1})\,\widehat{\rho}_{t_{0}}\,\widehat{\overline{E}%
}_{1}(t_{1})\,\widehat{E}_{2}(t_{2}))=0. \label{ch4}%
\end{equation}

Logical operations and relations are well defined on a family of consistent
histories. If $\Sigma_{1}=\Delta_{1}\cup\overline{\Delta}_{1}$ is the spectrum
of $\widehat{A}_{1}$ and $\Sigma_{2}=\Delta_{2}\cup\overline{\Delta}_{2}$ is
the spectrum of $\widehat{A}_{2}$, the elementary histories are represented in
$\Sigma_{1}\times\Sigma_{2}$ by the sets $\Delta_{1}\times\Delta_{2}$,
$\Delta_{1}\times\overline{\Delta}_{2}$, $\overline{\Delta}_{1}\times
\Delta_{2}$ and $\overline{\Delta}_{1}\times\overline{\Delta}_{2}$. All the
properties of the family are represented by the unions of these four sets. In
this way, the notions of \textit{conjunction}, \textit{disjunction} and
\textit{negation} are obtained. According to the approach of R. Omn\`{e}s
\cite{Om3}, \cite{Om2}, a property $a$ is said to \textit{imply} another
property $b$ of the same family if $\Pr(b|a)=\Pr(b$ and $a)/\Pr(a)=1$. The
conventional axioms of formal logic are satisfied with these definitions
\cite{Om2}. We notice that in this theory the implication relation is defined
trough a \textit{previous} definition of probability. This makes a big
difference with the usual approaches to quantum mechanics, where the logical
relations and operators on the properties have their own definition
independent of the probability, which is later on given by the Born rule. Our
own construction of the logical operators and relations, to be developed in
sections IV\ and V, will be established in a way which do not depend on the
definition of probability or on the state of the system.

R. Omn\`{e}s \cite{Om2} applied the theory of consistent histories to the case
of a $\frac{1}{2}$ spin system, prepared at the time $t_{0}$ in a pure state
with $S_{x}=+\frac{1}{2}\hbar$, represented by the vector $|x+\rangle$
($\widehat{\rho}_{t_{0}}=|x+\rangle\langle x+|$). The description of the
system include the two possible values of the spin along the $z$ axis
direction for the time $t_{2}$, which may be obtained by an Stern-Gerlach
experiment. With the simplifying assumption of the vanishing of the
Hamiltonian, he searched for the possibility to include in the description the
two possible values of the spin along a direction $\overline{n}_{1}$
($|\overline{n}_{1}|=1$) for a time $t_{1}$ after the preparation of the spin
along the positive direction of the $x$ axis, and before the time $t_{2}$
corresponding to the measurement along the $z$ axis ($t_{0}<t_{1}<t_{2}$). The
state vectors $|\overline{n}_{1}+\rangle$ and $|\overline{n}_{1}-\rangle$
correspond to the values $+\frac{1}{2}\hbar$ and $-\frac{1}{2}\hbar$ along the
direction $\overline{n}_{1}$. If the sufficient Gell-Mann and Hartle condition
(\ref{ch4}) is applied with $\widehat{E}_{1}(t_{1})=|\overline{n}_{1}%
+\rangle\langle\overline{n}_{1}+|$, $\widehat{\overline{E}}_{1}(t_{1}%
)=|\overline{n}_{1}-\rangle\langle\overline{n}_{1}-|$, $\widehat{\rho}%
_{0}=|x+\rangle\langle x+|$ and $\widehat{E}_{2}(t_{2})=|z+\rangle\langle
z+|$, two possibilities are obtained:

(i) A set of histories including the two possible spin values along the $x$
axis at time $t_{1}$, represented by the projectors $\widehat{E}%
_{1}=|x+\rangle\langle x+|$ and $\widehat{\overline{E}}_{1}=|x-\rangle\langle
x-|$, \textit{together with} the two possible spin values along the $z$ axis
at time $t_{2}$, represented by the projectors $\widehat{E}_{2}=|z+\rangle
\langle z+|$ and $\widehat{\overline{E}}_{2}=|z-\rangle\langle z-|$. Therefore
in this case $\overline{n}_{1}=(1,0,0)$.

(ii) A set of histories including the two possible spin values along the $z$
axis at time $t_{1}$, represented by the projectors $\widehat{E}%
_{1}=|z+\rangle\langle z+|$ and $\widehat{\overline{E}}_{1}=|z-\rangle\langle
z-|$, \textit{together with} the two possible spin values along the $z$ axis
at time $t_{2}$, represented by the projectors $\widehat{E}_{2}=|z+\rangle
\langle z+|$ and $\widehat{\overline{E}}_{2}=|z-\rangle\langle z-|$. In this
case $\overline{n}_{1}=(0,0,1)$.

More possible consistent histories are obtained using the necessary and
sufficient Griffiths condition given in Eq. (\ref{ch3}). If $\overline{n}_{0}$
is the preparation spin direction at the time $t_{0}$, and if $\overline
{n}_{2}$ is the spin direction at the time $t_{2}$, the equation
$(\overline{n}_{0}\times\overline{n}_{1})\cdot(\overline{n}_{1}\times
\overline{n}_{2})=0$ is obtained for the possible $\overline{n}_{1}$ spin
directions at the time $t_{1}$ (see reference \cite{Om4}, page 161). For
$\overline{n}_{0}=(1,0,0)$ and $\overline{n}_{2}=(0,0,1)$, representing the
$x$ and the $z$ directions, the direction $\overline{n}_{1}$ could be any
vector in the planes $xy$ or $yz$.

Up to now, we have considered well known facts of the theory of consistent
histories. We now analyze in more detail the consistent family obtained in the
case (i). Well defined probabilities can be obtained for all the members of
the family applying Eq. (\ref{ch2}). Let us consider the probability of the
spin to be $+\frac{1}{2}\hbar$ along the $x$ axis at time $t_{1}$ \textit{and}
to be $+\frac{1}{2}\hbar$ along the $z$ axis at time $t_{2}$,%
\[
\Pr((x+,t_{1});(z+,t_{2}))=Tr(\widehat{E}_{2}\widehat{E}_{1}\widehat{\rho
}_{t_{0}}\widehat{E}_{1}\widehat{E}_{2})=|\langle x+|z+\rangle|^{2}=\frac
{1}{2}.
\]

Provided that $t_{0}<t_{1}<t_{2}$, any value of $t_{1}$ gives a consistent
family of histories, and therefore a valid description of the spin system. As
the time $t_{1}$ can be chosen very close to the time $t_{2}$ we have%
\begin{equation}
\lim_{t_{1}\longrightarrow t_{2}}\Pr((x+,t_{1});(z+,t_{2}))=\frac{1}{2}.
\label{ch5}%
\end{equation}

The property giving simultaneously well defined values to different components
of the spin is forbidden in the universe of discourse of ordinary quantum
mechanics, due to the uncertainty principle (operators representing two
different components of the spin do not commute). Therefore the limit of Eq.
(\ref{ch5}) cannot be interpreted as the probability for the conjunction of
the spin values $x+$ and $z+$ at the single time $t_{2}$. The theory of
consistent histories is discontinuous in its property ascriptions for
different times.

In the following sections we present our own approach to the description of
time dependent properties of a quantum system. We are going to prove that in
our formalism only the case (ii) survives as a valid description for the two
times properties of the spin system. Only for this case, the limit
$t_{1}\longrightarrow t_{2}$ of Eq. (\ref{ch5}) can be interpreted as a single
time probability.

\section{Quantum logic, contexts and probabilities.}

We shall give in this section a brief summary of the logical structure of
quantum mechanics according to G. Birkhoff and J. von Neumann \cite{Bir},
which is one of the standard approaches to quantum logic \cite{Dalla}. This
summary is a preparation for our own approach to the problem of time dependent
properties, to be developed in section IV.

A Hilbert space $\mathcal{H}$ is associated with each isolated physical
system. Every quantum property $p$ is represented by a subspace $\mathcal{H}%
_{p}$ of the Hilbert space $\mathcal{H}$. For each subspace $\mathcal{H}_{p}$
there exists a projection operator $\widehat{\Pi}_{p}$ such that
$\mathcal{H}_{p}=\widehat{\Pi}_{p}\mathcal{H}$, and therefore a property $p$
can also be represented by the projector $\widehat{\Pi}_{p}$.

The \textit{implication relation} between two properties is \textit{defined}
by the inclusion of the corresponding subspaces ($p_{1}\Rightarrow p_{2}$ iff
$\mathcal{H}_{p_{1}}\subseteq\mathcal{H}_{p_{2}}$). The \textit{conjunction}
of two properties $p$ and $p^{\prime}$ is represented by the greatest lower
bound of the two corresponding subspaces $\mathcal{H}_{p}$ and $\mathcal{H}%
_{p^{\prime}}$ ($\mathcal{H}_{p\wedge p^{\prime}}=Inf(\mathcal{H}%
_{p},\mathcal{H}_{p^{\prime}})=\mathcal{H}_{p}\cap\mathcal{H}_{p^{\prime}}$),
while the \textit{disjunction} is the least upper bound ($\mathcal{H}_{p\vee
p^{\prime}}=Sup(\mathcal{H}_{p},\mathcal{H}_{p^{\prime}})$). Moreover, the
\textit{negation} of a property $p$ is represented by the orthogonal
complement of the corresponding subspace ($\mathcal{H}_{-p}=\mathcal{H}%
_{p}^{\perp}$).

Endowed with these implication, conjunction, disjunction and negation, the set
of all properties of a physical system is an orthocomplemented nondistributive
lattice. According to quantum theory, not all the properties can
simultaneously be considered in a description of a physical system. Different
descriptions involve different sets of properties.

Each possible description is called a \textit{context}, and it is defined
through a set of \textit{atomic or elementary properties} $p_{j}$ (where $j$
belongs to a set $\sigma$ of indexes). The properties $p_{j}$, represented by
projectors $\widehat{\Pi}_{j}$, are mutually exclusive and complete, i.e.%
\[
\widehat{\Pi}_{i}\widehat{\Pi}_{j}=\delta_{ij}\widehat{\Pi}_{i},\qquad
\sum_{i\in\sigma}\widehat{\Pi}_{i}=\widehat{I}.
\]

By disjunction of these elementary properties it is possible to generate all
the properties of the context, obtaining a \textit{distributive lattice}. Each
property $p$ of the context can be represented by a projector which is a sum
of \textit{some} of the projectors $\widehat{\Pi}_{i}$,%
\[
\widehat{\Pi}_{p}=\sum_{i\in\sigma_{p}}\widehat{\Pi}_{i},\qquad\sigma
_{p}\subset\sigma.
\]

A state of the system is represented by the statistical operator
$\widehat{\rho}$, which is self adjoint, positive and with trace equal to one.
In the state represented by $\widehat{\rho}$, the probability for each
property $p$ of a context is given by the Born rule%
\[
\Pr(p)=Tr(\widehat{\rho}\,\widehat{\Pi}_{p}).
\]

This rule gives a well defined probability (i.e. it is positive, normalized,
and satisfy the additivity property) when it is applied to the properties of a
given context.

Moreover, the probability of a property $b$ conditional to a property $a$,
\textit{defined} by%
\begin{equation}
\Pr(b|a)=\frac{\Pr(b\wedge a)}{\Pr(a)}, \label{d}%
\end{equation}
is also a well defined probability within a context. The conditional
probability can be used to give a statistical meaning to the implication
relation. It is not difficult to prove that a property $a$ implies a property
$b$ ($a\Rightarrow b$) \textit{if and only if} $\Pr(b|a)=1$ for \textit{all}
the states of the system.

We emphasize that the conjunction and the disjunction of the logical structure
of the quantum properties are obtained from the previous notion of
implication, defined by the inclusion of subspaces. On this lattice of
properties, the Born rule is applied to obtain the probabilities, which are
only meaningful for subsets of properties within a context. Properties
belonging to different contexts do not have simultaneous physical meaning.

In the next section we are going to present an extended notion of context to
deal with descriptions involving time dependent properties.

\section{Time translations and the lattice of time dependent properties.}

The time enters quantum theory through the Schr\"{o}dinger equation generating
the evolution of the state of an isolated system. The evolution of a vector of
the Hilbert space $\mathcal{H}$, representing a pure state, is given by%
\begin{equation}
|\varphi_{t^{\prime}}\rangle=\widehat{U}(t^{\prime},t)\,|\varphi_{t}%
\rangle,\qquad\widehat{U}(t^{\prime},t)=e^{-i\widehat{H}(t^{\prime}-t)/\hbar},
\label{A}%
\end{equation}
where $\widehat{H}$ is the Hamiltonian operator of the system.

Let us consider the physical system at the time $t$, in a pure state
represented by the vector $|\varphi_{t}\rangle$. Moreover, assume that
$|\varphi_{t}\rangle$ is an eigenvector with eigenvalue one of the projector
$\widehat{\Pi}_{p}$ corresponding to the property $p$%
\begin{equation}
\widehat{\Pi}_{p}|\varphi_{t}\rangle=|\varphi_{t}\rangle. \label{B}%
\end{equation}
We can say that the physical system \textit{has} the property $p$ at the time
$t$, because it has probability equal to one, according to the Born rule:%
\[
\Pr(p)=\langle\varphi_{t}|\widehat{\Pi}_{p}|\varphi_{t}\rangle=\langle
\varphi_{t}|\varphi_{t}\rangle=1.
\]

At a later time $t^{\prime}$, the time evolved state is represented by the
vector $|\varphi_{t^{\prime}}\rangle$ given by equation (\ref{A}). Using
equations (\ref{A}) and (\ref{B}) it is easy to prove that%
\[
\widehat{\Pi}_{p^{\prime}}|\varphi_{t^{\prime}}\rangle=|\varphi_{t^{\prime}%
}\rangle,\qquad\widehat{\Pi}_{p^{\prime}}=\widehat{U}(t^{\prime}%
,t)\widehat{\Pi}_{p}\widehat{U}^{-1}(t^{\prime},t).
\]

Therefore, if the system has the property represented by the projector
$\widehat{\Pi}_{p}$ at time $t$, it also has the property represented by the
projector $\widehat{\Pi}_{p^{\prime}}$ at time $t^{\prime}$. We have obtained
in this way a procedure for the \textit{time translation of properties}.

It is easily proved that the obtained relation between $p$ at time $t$ and
$p^{\prime}$ at time $t^{\prime}$ is transitive, reflexive and symmetric.
Therefore, it is an equivalence relation, that we shall indicate by the
expression $(\widehat{\Pi}_{p},t)\backsim(\widehat{\Pi}_{p^{\prime}}%
,t^{\prime})$.

The expression $(\widehat{\Pi}_{p},t)$ is a symbol indicating the
\textit{property }$p$\textit{ at time }$t$. We shall also indicate by
$[\widehat{\Pi}_{p},t]$ the \textit{class of properties equivalent to the
property }$p$\textit{ at time }$t$.

If a property $p$ at time $t$ is equivalent to $p^{\prime}$ at time
$t^{\prime}$, the Born rule gives for them the same probability,%
\begin{equation}
\Pr(\widehat{\Pi}_{p^{\prime}},t^{\prime})=Tr(\widehat{\rho}_{t^{\prime}%
}\widehat{\Pi}_{p^{\prime}})=Tr(\widehat{U}(t^{\prime},t)\widehat{\rho}%
_{t}\widehat{U}^{-1}(t^{\prime},t)\widehat{\Pi}_{p^{\prime}})=Tr(\widehat
{\rho}_{t}\widehat{\Pi}_{p})=\Pr(\widehat{\Pi}_{p},t), \label{B'}%
\end{equation}
and therefore a single probability is obtained for the properties of the same
class of equivalence. In physical terms, all the properties of a given class
are essentially the same property, as they can be obtained one from the other
trough time evolution.

The just considered time translation, and the implication of ordinary quantum
mechanics presented in the previous section, suggest that we \textit{define}
that the equivalence class $[\widehat{\Pi}_{1},t_{1}]$ \textit{implies} the
equivalence class $[\widehat{\Pi}_{2},t_{2}]$ if the representative elements
at a common time $t_{0}$ verify the implication of the usual formalism of
quantum mechanics, i.e.%
\[
\widehat{\Pi}_{1,0}\,\mathcal{H\subset}\widehat{\Pi}_{2,0}\mathcal{H}%
,\qquad\widehat{\Pi}_{1,0}\equiv\widehat{U}(t_{0},t_{1})\widehat{\Pi}%
_{1}\widehat{U}^{-1}(t_{0},t_{1}),\qquad\widehat{\Pi}_{2,0}\equiv\,\widehat
{U}(t_{0},t_{2})\widehat{\Pi}_{2}\widehat{U}^{-1}(t_{0},t_{2})\,\mathcal{H}.
\]

It is not difficult to prove that if two projectors $\widehat{\Pi}_{1}$ and
$\widehat{\Pi}_{2}$ verify this condition for a given time $t_{0}$, they
verify the condition for all possible values of $t_{0}$. The implication
relation is transitive, reflexive and antisymmetric, and therefore it is a
well defined \textit{order relation} on the equivalence classes.

Having defined an order relation on the equivalence classes, the
\textit{conjunction} (\textit{disjunction}) of two classes $[\widehat{\Pi},t]$
and $[\widehat{\Pi}^{\prime},t^{\prime}]$ can be obtained as the greatest
lower (least upper) bound, i.e.%
\begin{equation}
\lbrack\widehat{\Pi},t]\wedge\lbrack\widehat{\Pi}^{\prime},t^{\prime
}]=Inf\{[\widehat{\Pi},t];[\widehat{\Pi}^{\prime},t^{\prime}]\}=[\lim
_{n\rightarrow\infty}(\widehat{\Pi}_{0}\widehat{\Pi}_{0}^{\prime})^{n},t_{0}],
\label{C}%
\end{equation}%
\begin{equation}
\lbrack\widehat{\Pi},t]\vee\lbrack\widehat{\Pi}^{\prime},t^{\prime
}]=Sup\{[\widehat{\Pi},t];[\widehat{\Pi}^{\prime},t^{\prime}]\}=[(\widehat
{I}-\lim_{n\rightarrow\infty}\{(\widehat{I}-\widehat{\Pi}_{0})(\widehat
{I}-\widehat{\Pi}_{0}^{\prime})\}^{n}),t_{0}], \label{D}%
\end{equation}
where $\widehat{\Pi}_{0}=\widehat{U}(t_{0},t)\,\widehat{\Pi}\,\widehat{U}%
^{-1}(t_{0},t)$ and $\widehat{\Pi}_{0}^{\prime}=\widehat{U}(t_{0},t^{\prime
})\,\widehat{\Pi}^{\prime}\,\widehat{U}^{-1}(t_{0},t^{\prime})$ are the
translations of the properties $\widehat{\Pi}$ and $\widehat{\Pi}^{\prime}$ to
the time $t_{0}$. The projectors $\lim_{n\rightarrow\infty}(\widehat{\Pi}%
_{0}\widehat{\Pi}_{0}^{\prime})^{n}$ and $(\widehat{I}-\lim_{n\rightarrow
\infty}\{(\widehat{I}-\widehat{\Pi}_{0})(\widehat{I}-\widehat{\Pi}_{0}%
^{\prime})\}^{n})$ generate the greatest lower and the least upper bounds of
the subspaces generated by $\widehat{\Pi}_{0}$ and $\widehat{\Pi}_{0}^{\prime
}$ \cite{Mit}.

The \textit{negation} of an equivalence class $[\widehat{\Pi},t]$ is defined
by%
\[
\overline{\lbrack\widehat{\Pi},t]}=[\widehat{\overline{\Pi}},t]=[(\widehat
{I}-\widehat{\Pi}),t].
\]

With the implication, disjunction, conjunction and negation previously
obtained, the set of equivalent classes has the structure of an
orthocomplemented nondistributive lattice.

\section{The generalized contexts.}

The usual concept of context was briefly reviewed in section III as a subset
of all possible simultaneous properties which can be organized as a meaningful
description of a quantum system at a given time, and endowed with a boolean
logic with well defined probabilities.

The definitions and notations given in the previous section will be useful to
our purpose of representing valid \textit{descriptions involving properties at
different times}, which we are going to call \textit{generalized contexts}. In
what follows, we shall only consider descriptions involving properties at two
times $t_{1}$ and $t_{2}$, but our formalism has an immediate extension to
cases involving more than two times.

Let us consider a context of properties at time $t_{1}$, generated by atomic
properties $p_{j}^{(1)}$ represented by projectors $\widehat{\Pi}_{j}^{(1)}$
verifying%
\begin{equation}
\widehat{\Pi}_{i}^{(1)}\widehat{\Pi}_{j}^{(1)}=\delta_{ij}\,\widehat{\Pi}%
_{i}^{(1)},\qquad\sum_{j\in\sigma^{(1)}}\widehat{\Pi}_{j}^{(1)}=\widehat
{I},\qquad i,j\in\sigma^{(1)}. \label{E}%
\end{equation}

Let us also consider a context of properties at time $t_{2}$, generated by
atomic properties $p_{\mu}^{(2)}$ represented by projectors $\widehat{\Pi
}_{\mu}^{(2)}$ verifying%
\begin{equation}
\widehat{\Pi}_{\mu}^{(2)}\widehat{\Pi}_{\nu}^{(2)}=\delta_{\mu\nu}%
\,\widehat{\Pi}_{\mu}^{(2)},\qquad\sum_{\mu\in\sigma^{(2)}}\widehat{\Pi}_{\mu
}^{(2)}=\widehat{I},\qquad\mu,\nu\in\sigma^{(2)}. \label{F}%
\end{equation}

We wish to represent with our formalism a universe of discourse capable to
incorporate expressions like "the property $p_{j}^{(1)}$ at time $t_{1}$
\textit{and} the property $p_{\mu}^{(2)}$ at time $t_{2}$". With this purpose,
we note that the properties associated to different times $t_{1}$ and $t_{2}$
can be translated to a common time $t_{0}$, by using the equivalence relations
previously defined%
\begin{align}
(\widehat{\Pi}_{i}^{(1)},t_{1})  &  \backsim(\widehat{\Pi}_{i}^{(1,0)}%
,t_{0}),\qquad\widehat{\Pi}_{i}^{(1,0)}\equiv\widehat{U}(t_{0},t_{1}%
)\widehat{\Pi}_{i}^{(1)}\widehat{U}^{-1}(t_{0},t_{1}),\nonumber\\
(\widehat{\Pi}_{\mu}^{(2)},t_{2})  &  \backsim(\widehat{\Pi}_{\mu}%
^{(2,0)},t_{0}),\qquad\widehat{\Pi}_{\mu}^{(2,0)}\equiv\widehat{U}(t_{0}%
,t_{2})\widehat{\Pi}_{\mu}^{(2)}\widehat{U}^{-1}(t_{0},t_{2}). \label{defi}%
\end{align}

The conjunction of the equivalence classes $[\widehat{\Pi}_{i}^{(1)},t_{1}]$
and $[\widehat{\Pi}_{\mu}^{(2)},t_{2}]$ can be obtained applying Eq.
(\ref{C})
\[
\lbrack\widehat{\Pi}_{i}^{(1)},t_{1}]\wedge\lbrack\widehat{\Pi}_{\mu}%
^{(2)},t_{2}]=[\widehat{\Pi}_{i}^{(1,0)},t_{0}]\wedge\lbrack\widehat{\Pi}%
_{\mu}^{(2,0)},t_{0}]=[\lim_{n\rightarrow\infty}(\widehat{\Pi}_{i}%
^{(1,0)}\widehat{\Pi}_{\mu}^{(2,0)})^{n},t_{0}].
\]

The conjunction of the classes with representative elements $\widehat{\Pi}%
_{i}^{(1)}$ at $t_{1}$ and $\widehat{\Pi}_{\mu}^{(2)}$ at $t_{2}$, is also the
conjunction of the classes with representative elements $\widehat{\Pi}%
_{i}^{(1,0)}$ and $\widehat{\Pi}_{\mu}^{(2,0)}$ at the common time $t_{0}$.
Moreover, the conjunction is a class with the representative element
$\lim_{n\rightarrow\infty}(\widehat{\Pi}_{i}^{(1,0)}\widehat{\Pi}_{\mu
}^{(2,0)})^{n}$ at the time $t_{0}$. The conjunction of properties at the same
time is already defined in quantum mechanics, \textit{for the particular case
in which they are represented by commuting projectors}. The usual quantum
theory do not give any meaning to the conjunction of simultaneous properties
represented by non commuting operators.

To make contact with the usual formalism of quantum theory, it seems natural
to consider quantum descriptions of a system, involving the properties
generated by the projectors $\widehat{\Pi}_{i}^{(1)}$ at the time $t_{1}$ and
$\widehat{\Pi}_{\mu}^{(2)}$ at the time $t_{2}$, \textit{only for the cases}
in which the projectors $\widehat{\Pi}_{i}^{(1)}$ and $\widehat{\Pi}_{\mu
}^{(2)}$ commute when translated to a common time $t_{0}$, i.e.%
\begin{equation}
\widehat{\Pi}_{i}^{(1,0)}\widehat{\Pi}_{\mu}^{(2,0)}-\widehat{\Pi}_{\mu
}^{(2,0)}\widehat{\Pi}_{i}^{(1,0)}=0 \label{F'}%
\end{equation}

If this is the case, we have $\lim_{n\rightarrow\infty}(\widehat{\Pi}%
_{i}^{(1,0)}\widehat{\Pi}_{\mu}^{(2,0)})^{n}=\widehat{\Pi}_{i}^{(1,0)}%
\widehat{\Pi}_{\mu}^{(2,0)}$, and for the equivalence class of composed
properties $h_{i\mu}$, representing "the property $p_{j}^{(1)}$ at time
$t_{1}$ \textit{and} the property $p_{\mu}^{(2)}$ at time $t_{2}$" we obtain%
\[
h_{i\mu}=[\widehat{\Pi}_{i}^{(1)},t_{1}]\wedge\lbrack\widehat{\Pi}_{\mu}%
^{(2)},t_{2}]=[\widehat{\Pi}_{i}^{(1,0)}\widehat{\Pi}_{\mu}^{(2,0)}%
,t_{0}]=[\widehat{\Pi}_{i\mu}^{(0)},t_{0}],\qquad\widehat{\Pi}_{i\mu}%
^{(0)}\equiv\widehat{\Pi}_{i}^{(1,0)}\widehat{\Pi}_{\mu}^{(2,0)}.
\]

As we can see, the conjunction of properties at different times $t_{1}$ and
$t_{2}$ is equivalent to a single property represented by the projector
$\widehat{\Pi}_{i\mu}^{(0)}$ at the single time $t_{0}$.

If the different contexts at times $t_{1}$ and $t_{2}$ produce commuting
projectors $\widehat{\Pi}_{i}^{(1,0)}$ and $\widehat{\Pi}_{\mu}^{(2,0)}$ at
the common time $t_{0}$, it is easy to prove that%
\begin{equation}
\widehat{\Pi}_{i\mu}^{(0)}\widehat{\Pi}_{j\nu}^{(0)}=\delta_{ij}\delta_{\mu
\nu}\widehat{\Pi}_{i\mu}^{(0)},\qquad\sum_{i\mu}\widehat{\Pi}_{i\mu}%
^{(0)}=\widehat{I}. \label{G}%
\end{equation}

Therefore, we realize that the composed properties $h_{i\mu}$, represented at
the time $t_{0}$ by the complete and exclusive set of projectors $\widehat
{\Pi}_{i\mu}^{(0)}$, can be interpreted as the atomic properties generating a
usual context in the sense already described in the previous section. More
general properties are obtained from the atomic ones using the disjunction
operation defined in Eq. (\ref{D}). For example, taking into account the
commutation relation (\ref{F'}), we obtain%
\[
h_{i\mu}\vee h_{j\nu}=[\widehat{\Pi}_{i\mu}^{(0)}+\widehat{\Pi}_{j\nu}%
^{(0)},t_{0}].
\]

More generally, we can represent the property $p_{j}^{(1)}$ at time $t_{1}$
\textit{and} the property $p_{\mu}^{(2)}$ at time $t_{2}$, with $j$ and $\mu$
having any value in the subsets $\Delta^{(1)}\subset\sigma^{(1)}$ and
$\Delta^{(2)}\subset\sigma^{(2)}$, in the form%
\begin{equation}
h_{\Delta^{(1)},\Delta^{(2)}}=[\sum_{i\in\Delta^{(1)}}\sum_{\mu\in\Delta
^{(2)}}\widehat{\Pi}_{i\mu}^{(0)},t_{0}] \label{H}%
\end{equation}

As a consequence of Eqs. (\ref{G}), the set of properties obtained in this way
is an orthocomplemented and distributive lattice.

As we proved in Eq. (\ref{B'}), the Born rule defines a single probability to
all elements of an equivalence class. If the state of the system at time
$t_{0}$ is represented by $\widehat{\rho}_{t_{0}}$, the probability of the
class of properties $h_{\Delta^{(1)},\Delta^{(2)}}$ has the following
expression%
\begin{equation}
\Pr(h_{\Delta^{(1)},\Delta^{(2)}})=\sum_{i\in\Delta^{(1)}}\sum_{\mu\in
\Delta^{(2)}}Tr(\widehat{\rho}_{t_{0}}\widehat{\Pi}_{i\mu}^{(0)}). \label{I}%
\end{equation}

As we already mentioned in the previous section, a description of a physical
system should not involve properties belonging to different contexts. As a
natural extension of the notion of context, we \textit{postulate} that a
description of a physical system involving properties at two different times
$t_{1}$ and $t_{2}$ is valid if these properties are represented by commuting
projectors when they are translated to a single time $t_{0}$. We will call
\textit{generalized context} to each of these valid descriptions. On each
generalized context, the probabilities given by the Born rule are well defined
(i.e. they are positive, normalized and additive), and therefore they may have
a meaning in terms of frequencies.

In summary, our formalism is based on the notion of time translation, allowing
to transform the properties at a sequence of different times into properties
at a single common time. A usual context of properties is first considered for
each time of the sequence. If the projectors representing the atomic
properties of each context commute when they are translated to a common time,
the contexts at different times can be organized forming a generalized context
of properties. A generalized context of properties is a distributive and
orthocomplemented lattice, a boolean logic with well defined implication,
negation, conjunction and disjunction. This logic can be used for speaking and
reasoning about the selected properties of the system at different times. Well
defined probabilities on the elements of the lattice of properties are
obtained using the well known Born rule.

\section{Comparison of the generalized contexts with the sets of consistent
histories.}

In our opinion, the generalized contexts seem to be a natural generalization
of the usual contexts of quantum mechanics. They are suitable to deal with the
logic of properties at different times. But to be a "natural generalization"
may have no any scientific value, and perhaps may only reflects our confidence
in the usual form of quantum theory. Therefore, it is necessary to compare our
new approach with the theory of consistent histories, designed to deal with
the same kind of problems, and also to apply the new formalism to physically
relevant situations. The main relation between both theories is given by the
following Theorem:

\textit{A generalized context obtained with our formalism, is also a
consistent set of histories, with the same probabilities}.

We give the proof for a generalized context with two times. The probability
for the property $p_{j}^{(1)}$ at time $t_{1}$ \textit{and} the property
$p_{\mu}^{(2)}$ at time $t_{2}$, is given by Eq. (\ref{I})%
\begin{align*}
\Pr((p_{j}^{(1)},t_{1})\wedge(p_{\mu}^{(2)},t_{2}))  &  =Tr(\widehat{\rho
}_{t_{0}}\widehat{\Pi}_{j\mu}^{(0)})=Tr(\widehat{\rho}_{t_{0}}\widehat{\Pi
}_{j}^{(1,0)}\widehat{\Pi}_{\mu}^{(2,0)})\\
&  =Tr(\widehat{\Pi}_{\mu}^{(2,0)}\widehat{\Pi}_{j}^{(1,0)}\widehat{\rho
}_{t_{0}}\widehat{\Pi}_{j}^{(1,0)}\widehat{\Pi}_{\mu}^{(2,0)}),
\end{align*}
where the last equality is a consequence of the commutation relation
(\ref{F'}) and the cyclic permutation of the operators in the trace. Taking
into account the definitions (\ref{defi}), we obtain for the probability the
same expression which is obtained with Eq. (\ref{ch2}) for consistent
histories. Moreover, the consistency conditions $Tr(\widehat{C}_{a}%
\,\widehat{\rho}_{t_{0}}\,\widehat{C}_{b}^{\dag})=0$, for $a\neq b$, are
satisfied due to the commutation relations (\ref{F'}). A simple generalization
of this proof is obtained for a generalized context with $n$ times. In simple
words, the meaning of this theorem is that our formalism put more restrictions
than the theory of consistent histories on the number of valid descriptions of
a physical system.

A search of the sets of consistent histories which are forbidden by our
formalism, and their physical relevance, is unavoidable.

We can analyze with our formalism the spin $\frac{1}{2}$ system, already
described using the theory of consistent histories at the end of section II.
Once again we consider a description including the two possible values of the
spin along the $z$ axis for the time $t_{2}$, and we ask which properties can
also be considered at the time $t_{1}$ ($t_{0}<t_{1}<t_{2}$), in such a way
that they are compatible with the properties chosen at the time $t_{2}$.

The atomic properties for the time $t_{2}$ are represented by the projectors
$\widehat{E}_{z+}=|z+\rangle\langle z+|$ and $\widehat{E}_{z-}=|z-\rangle
\langle z-|$, while the atomic properties at $t_{1}$ are represented by
$\widehat{E}_{\overline{n}_{1}+}=|\overline{n}_{1}+\rangle\langle\overline
{n}_{1}+|$ and $\widehat{E}_{\overline{n}_{1}-}=|\overline{n}_{1}%
-\rangle\langle\overline{n}_{1}-|$, for an unknown direction $\overline{n}%
_{1}$ of the spin. These projectors are invariant under time translations, due
to the vanishing of the Hamiltonian. Therefore, they are invariant when
translated to any common time. We may choose this common time as $t_{0}$
($t_{0}<t_{1}<t_{2}$), where the initial state $\widehat{\rho}_{t_{0}}$ is
given. If the commutation conditions (\ref{F'}) are satisfied, we should have
$\widehat{E}_{\overline{n}_{1}\pm}\widehat{E}_{z\pm}-\widehat{E}_{z\pm
}\widehat{E}_{\overline{n}_{1}\pm}=\widehat{E}_{\overline{n}_{1}\mp}%
\widehat{E}_{z\pm}-\widehat{E}_{z\pm}\widehat{E}_{\overline{n}_{1}\mp}=0$,
which gives the $z$ direction ($\overline{n}_{1}=(0,0,1)$) as the only
possibility. The $z$ components of the spin at time $t_{1}$ is the only choice
compatible with the $z$ components at the time $t_{2}$, and it corresponds to
the case (ii) obtained with consistent histories in section II. Moreover, this
is the only possible choice \textit{for any initial state} $\widehat{\rho
}_{t_{0}}$.

The case (i) for the Gelmann and Hartle condition, and all the possibilities
for the Griffiths condition are ruled out by our formalism of generalized
contexts. Only the case (ii), which is time continuous with respect to
property ascriptions, remains.

It is also necessary to verify if the postulated compatibility condition for
time translated properties is successful to give a good description of well
established physical processes.

We only mention here what are the results for the well known double slit
experiment. R. Omn\`{e}s \cite{Omn3} proved that with no measurement
instruments there is no place for a set of consistent histories including in
its universe of discourse through which slit passed the particle before
reaching a zone in front of the slits. Therefore, as a consequence of the
theorem given at the beginning of this section, there is also no room for such
a description with our generalized contexts. For the case of the double slit
with a measurement instruments recording through which slit passed the
particle, and another instrument recording the particle in different zones of
a plane in front of the double slit, we found with our approach the existence
of a generalized context suitable for the description of the registration of
the instruments (but not of the particle positions). As a consequence, the
theorem of the beginning of this section can be used to deduce that such a
description has also a place in a set of consistent histories. These are
preliminary results which will be included in a forthcoming paper.

The version of the theory of consistent histories given by R. Omn\`{e}s
\cite{Om3}, \cite{Om2}, \cite{Om4} emphasizes its role as a logical
construction, i.e. as a tool for obtaining valid descriptions and reasonings
about properties of the system. As this is also the case in our formalism, it
is interesting to compare both logical structures.

As we briefly summarized in section II, in the theory of consistent histories
there are ordinary contexts on each time of the sequence. The conjunction,
disjunction and negation of properties at different times are defined through
the union, intersection and complement of the corresponding spectrums, as
shown for the two times case in the paragraph below Eq. (\ref{ch4}). In this
theory, a history $a$ implies a history $b$ when $\Pr(b|a)\equiv\Pr(b\wedge
a)/\Pr(a)=1$. As the probabilities depend on the state, the implication of the
theory is also state dependent. If the set of histories verify the state
dependent consistency conditions given by Eqs. (\ref{SC}), (\ref{ch3}) or
(\ref{ch4}), it is named a set of consistent histories, and within this set
the conventional axioms of formal logic are satisfied. Therefore, the possible
universes of discourse provided by this theory have a very special
entanglement with the state of the system.

This situation is not entirely satisfactory, because in the usual axiomatic
theories of quantum mechanics the state is considered as a functional on the
space of observables, and it appears after these observables in a somehow
subordinate position. The importance of the notion of state functionals acting
on a previously defined space of observables was stressed by one of us in
references \cite{RL1} and \cite{RL2}. In our approach of sections IV and V,
the logical structure of the properties is an orthocomplemented lattice
defined independently of the state of the system and of the probability
definition. Moreover, the conditions to have a generalized context are
commutation relations, also state independent (see the condition given by Eq.
(\ref{F'}) for the two times case). Probability is later on introduced on the
already constructed logical structure, trough the usual Born rule.

\section{Conclusions.}

We have introduced in this paper a formalism suitable to deal with
descriptions and reasonings about physical systems involving quantum
properties at different times. The dynamic generated by the Schr\"{o}dinger
equation provides a natural definition for the time translation of quantum
properties. Time translations generate a partition in equivalence classes of
the set of properties and times. From a physical point of view, properties at
different times which are connected by a time translation are essentially the
same property, on which the Born rule gives the same probability value.

Time translation also provide the possibility to define an implication between
classes. We used this implication to obtain through infimum and supremum the
definitions of conjunction and disjunctions of classes. The orthogonal
complement of Hilbert spaces is immediately generalized to obtain the negation
of a class. In this way we construct a non distributive orthocomplemented
lattice of classes of properties and times, and we obtain what in our opinion
is a natural extension of the logical structure of quantum mechanics given by
Birkhoff and J. von Neumann \cite{Bir}, one of the standard approaches to
quantum logic.

As the lattice is non distributive, the Born rule do not provide a well
defined probability on the whole of it. Therefore, we extended the usual
notion of context to the notion of generalized context, which is a subset of
the whole set of classes, organized in a distributive and orthocomplemented
lattice. On each generalized context, the Born rule provides a well defined
probability. A generalized context is a boolean logic which can be used for
speaking and reasoning about properties of the system at different times. It
is interesting to note that our formalism allows to define the logic of
quantum properties without referring to any state of the system under consideration.

Our approach impose more restrictions than the theory of consistent histories
on the possible valid descriptions of a physical system. For a spin system, we
proved that our more restrictive conditions eliminate the sets of consistent
histories which do not satisfy time continuity for the property ascriptions.
This continuity is in our opinion a desirable property, and it is a direct
consequence of the fact that our formalism only allows quantum properties
represented by commuting projectors when translated to a common time.

We also obtained good preliminary results with our approach describing the
double slit experiment with and without measurement instruments detecting the
particle passing trough the slits. This open the possibility to apply our
formalism to the description of the measurement process and to the classical
limit, and moreover to explore in this framework the role of the environment
induced decoherence. The work in this direction is in progress.

\end{document}